\documentclass[acmsmall]{acmart}
\usepackage[]{graphicx}
\usepackage{subfigure}
\usepackage{algorithm}
\usepackage[noend]{algpseudocode} 
\usepackage{multirow}

\AtBeginDocument{%
  \providecommand\BibTeX{{%
    \normalfont B\kern-0.5em{\scshape i\kern-0.25em b}\kern-0.8em\TeX}}}

\begin{document}

\title{Modeling and Mitigating Online Misinformation: a Suggested Blockchain Approach}

\author{Tolga Yilmaz}
\email{tolga.yilmaz@bilkent.edu.tr}
\orcid{1234-5678-9012}
\author{Özgür Ulusoy}
\email{oulusoy@cs.bilkent.edu.tr}
\affiliation{%
  \institution{Bilkent University}
  \streetaddress{Bilkent, Ankara}
  \city{Ankara}
  \country{Turkey}
  \postcode{06800}
}

\renewcommand{\shortauthors}{Yilmaz and Ulusoy}

\begin{abstract}
Misinformation propagation in online social networks has become an increasingly challenging problem. Although many studies exist to solve the problem computationally, a permanent and robust solution is yet to be discovered. In this study, we propose and demonstrate the effectiveness of a blockchain-machine learning hybrid approach for addressing the issue of misinformation in a crowdsourced environment. First, we motivate the use of blockchain for this problem by finding the crucial parts contributing to the dissemination of misinformation and how blockchain can be useful, respectively. Second, we propose a method that combines the wisdom of the crowd with a behavioral classifier to classify the news stories in terms of their truthfulness while reducing the effects of the actions performed by malicious users. We conduct experiments and simulations under different scenarios and attacks to assess the performance of this approach. Finally, we provide a case study involving a comparison with an existing approach using Twitter Birdwatch data. Our results suggest that this solution holds promise and warrants further investigation.

\end{abstract}



\maketitle

\section{Introduction}
The proliferation of misinformation has become a major concern in recent years. With the rise of social media and the ease of access to information, it has become increasingly difficult to distinguish between factual and false information. This has led to the spread of false narratives and even conspiracy theories in politics \cite{allcott2017}, vital health-related issues such as pandemics \cite{apuke2021fake}, disasters \cite{gupta2013faking}, energy \cite{ho2022fake}, terrorism \cite{piazza2022fake} and armed conflicts \cite{khaldarova2020fake}, which can have serious consequences for individuals and society as a whole.

To combat the spread of misinformation, some organizations have attempted to raise awareness about "fake news" by manually curating content. Duke Reporter's Lab\footnote{https://reporterslab.org/fact-checking/} maintains a database of fact-checking websites around the world. However, this approach has its limitations. These organizations are often centralized, meaning that they are controlled by a small group of people. This can make it difficult for them to guarantee transparency and prevent manipulation, such as cherry-picking in favor of a particular party \cite{shin2017partisan, babaei2021analyzing, draws2022effects}. In addition, "fact-checking" is labor-intensive, and the curators may not be able to keep up with emerging news stories or cover the entire news ecosystem. By the time a human curator checks the content, it would already propagate to many people \cite{thorson2016belief}. \cite{guo2022survey} surveys the recent efforts on automated fact-checking to accelerate the process.

News literacy education is also important to teach individuals how to critically evaluate the information they encounter and to understand the various biases and motivations that can influence the production and dissemination of news. This can help individuals to become more discerning consumers of information and to avoid being swayed by false narratives \cite{higdon2020anatomy}. As in any form of education, news literacy education can take time to have an effect, and it may not reach everyone who needs it. 

Computational approaches focus on the definition of the misinformation concept \cite{kumar2018}, its propagation mechanisms \cite{vosoughi2018spread, DelVicario554}, and detection and preventing its propagation \cite{Qazvinian2011, shrivastava2020defensive, li2020characterizing, yilmaz2022misinformation}. Recent studies on this field iterate over various types of textual, behavioral, and media aspects to identify misinformation using machine learning as explained by \cite{hu2022deep}, \cite{zhou2020survey}, and \cite{zhang2020overview}. 
 Computational efforts should also take the concepts of decentralization, transparency, and objectivity into account and should have demonstrative aspects against tampering for any outside agenda \cite{mohseni2019open, seo2019trust, nguyen2018believe}.

It has been reported that large social media companies use techniques to detect fake news and troll accounts \cite{wingfield2016google}, which are intentionally misleading or disruptive online postings. However, despite such efforts, these platforms continue to be sources of misinformation \cite{kumar2018} as ongoing research on the spread of misinformation on these platforms highlights the ongoing issue \cite{zhou2020survey}. In addition, these platforms are commercially motivated and centrally controlled. This means that they may not always be able to provide transparency and objectivity in the way that they store, distribute, and present content and interactions \cite{bartley2021auditing, bandy2021curating}.

In this work, we approach the news-sharing notion as an issue of trust. As in any transaction between two parties, the news delivery process between the curator and the news-reader is established on some notion of trust \cite{fletcher2017}, and can be investigated from a trust management point of view. People should be able to find reputable news sources they can trust, while news curators are rewarded in parallel with the quality of their work and establish a form of reputation. However, the research on trust management systems that are engineered and deployed for various tasks shows that these systems can also be exploited with various types of attacks \cite{Hoffman:2009:SAD:1592451.1592452,wang2014,ruan2016survey}. Nonetheless, these systems deliver resilient enough functions that we incorporate into our lives, such as e-shopping \cite{HENDRIKX2015184}. 

A trust-based platform that can incorporate this kind of transparency and that can prove the validity of transactions with a transaction-based reward mechanism is a blockchain. A blockchain is a distributed database to deliver specific functionality while ensuring transaction correctness. In this work, we explore a blockchain-based crowdsourcing system using the Ethereum blockchain and smart contracts as a proof-of-concept solution where people create and vote on news stories. We employ a reward-based reputation system that encourages people to post correct stories and identify the truthfulness of stories correctly by voting on them. A blockchain-based solution practically ensures that the transactions are not altered or in the custody of a single body while providing incentives in the form of rewards for user traction. However, we show that, as in any reputation system, this system is also subject to various forms of attacks involving malicious users and should not be presented as a silver bullet but rather needs to be supported by various mechanisms for increased resilience and robustness. To this purpose, we reinforce the system with a machine learning architecture that learns the malicious actions of the users and joins the final decision of the crowd on the truthfulness of news stories. We simulate various types of attack forms as well as provide a case study based on Twitter Birdwatch. The experiments yield promising results and encourage further studies, which we specify in the paper.

To summarize, our contributions are as follows:
\begin{itemize}
    \item We identify the critical factors in deception by referencing the Interpersonal Deception Theory and provide a model for misinformation in online social networks. We describe how blockchains are one of the candidates that can provide a better solution.
    \item We describe and implement a crowdsourcing mechanism on the blockchain, specifically on Ethereum, using smart contracts towards identifying true and false stories.
    \item We add a machine learning system that takes part in identifying the malicious actions in the crowdsourcing mechanism and the final decision on the truthfulness of stories. 
    \item We train and test the system under various types of attacks and provide the results. 
    \item We provide a case study on Twitter Birdwatch data and compare our results with another study on the same data set.

\end{itemize}

The paper is structured as follows. In Section 2, we summarize the related work. Section 3 gives our interpretation of how misinformation occurs on social networks. Section 4 describes our methodology. Section 5 provides the details of our experimental evaluation. In Section 6, we discuss the effectiveness of the system and how it can be improved to address the misinformation problem with future research directions. Finally, in Section 7, we summarize our work in this chapter.

\section{Related Work}

We divide the research on misinformation into three main parts. First, we list the literature on how misinformation is viewed and modeled. Then, we give a short review of the studies on the detection and containment of misinformation. We finally provide an overview of more specific studies involving crowdsourcing and blockchains.

To classify misinformation, Kumar et al. \cite{kumar2018} provide an analysis from a "false information" perspective. According to their categorization, false information can be classified into \textit{intent} and \textit{knowledge}, which are further classified into misinformation, disinformation, opinion-based, and fact-based, respectively. Rubin et al. \cite{Rubin2015} define three types of fakes: serious fabrications, large-scale hoaxes, and humorous fakes. There exist recent studies that have similar classifications \cite{wu2019misinformation, zannettou2019web} and ones with from an intent perspective \cite{zhang2020overview} regarding the types of misinformation or fake news.

Some studies investigate the spread nature of misinformation on social networks. Jin et al. \cite{Jin2016} establish a method that utilizes opposing views to detect false information. Ruchansky et al. \cite{Ruchansky2017} utilize three characteristics of fake news; the article itself, the user behavior on the article, and the community that spreads it. They build a classifier of three modules, the first of which uses the first two characteristics with a Recurrent Neural Network. The second module is learning the third characteristic, and the third module combines the results. Wu et al. \cite{Wu2018} use a method that represents users with network structures as embeddings and Long Short Term Memory (LSTM) to learn the propagation structure in the network to classify misinformation. Ma et al. \cite{ma2017} represent propagation structures as trees and use a technique to learn these structures to identify rumors. 

Budak et al. \cite{Budak2011} define the misinformation limitation problem as limiting the spread of an adversarial campaign with a good one. Gupta et al. \cite{gupta2014tweetcred} describe a system that can classify tweets based on their credibility after collecting user feedback and learning a Support Vector Machine based classifier from them.  Nguyen et al. \cite{NGUYEN20132133} define a concept of \textit{node protectors}, which is the smallest subset of nodes when contained, which will prevent misinformation from spreading. Farajtabar et al. \cite{Farajtabar2017} define the misinformation spread as a reinforcement learning problem by representing states, mitigation actions, and reward functions to learn. Yilmaz et al. \cite{yilmaz2022misinformation} propose a model for misinformation propagation in social networks conforming to a game-theoretic model and also suggest a deep reinforcement learning-based game in which an agent learns to spread misinformation while the other opposes its propagation. It should be noted that the studies on containing misinformation have been intertwined with the ones on \textit{influence maximization}. Li et al. \cite{li2022survey} compile a recent and comprehensive survey on influence maximization using deep learning methods. Comprehensive reviews of recent misinformation detection and containment techniques are given in \cite{sharma2019combating, zhang2020overview, zhou2020survey}. 

Crowdsourcing can be an effective way to identify true and fake news through the use of collective intelligence. By gathering the perspectives and expertise of a diverse group of individuals, it is possible to assess the veracity of news stories and determine whether they are genuine or fabricated.

A considerable number of studies in the computer science literature have demonstrated the effectiveness of crowdsourcing in identifying true and fake news. In one of these studies, Kittur et al. \cite{kittur2008crowdsourcing} show that crowdsourced fact-checking can be more accurate than individual fact-checking, as the collective wisdom of the crowd can often outweigh the biases and limitations of any one individual.

Chen et al. \cite{chen2022implementation} propose an entropy-based mechanism for a quorum-based fake-news prevention system that uses expert voting on fake news candidates, with Hyperledger as the persistence layer. Avelino and Rocha \cite{avelino2022blockproof} suggest BlockProof for verifying the authenticity and integrity of web content using a blockchain solution. Their experiments mainly focus on the feasibility of the average response times of the system rather than how and how well the system identifies fake news. In \cite{sengupta2021problock}, Sengupta et al.  propose ProBlock, which is a secure voting system with an emphasis on reviewer privacy.

Pennycook and Rand \cite{pennycook2019fighting} measure the effectiveness of crowdsourcing to identify credible news sources. Denaux et al. \cite{denaux2020towards} work on identifying the credibility of reviews to increase the reliability of crowdsourcing tasks. Soprano et al. \cite{soprano2021many} provide an in-depth, multidimensional analysis of crowdsourcing for misinformation assessment, denoting the effectiveness of crowdsourcing for this problem. Teixeira et al. \cite{teixeira2020new} explore the preliminaries of a blockchain-based journalism system.

Shan et al. introduce Poligraph \cite{shan2021poligraph} which aims to provide a human review-machine learning combined approach for expert review on blockchain and focuses on latency and byzantine fault tolerance without putting the accuracy of the machine learning approach forward, where our work is reputation related and designed for the use of general public; hence the problem requires solving issues such as the attacks under uncertainty of user intent. A similar work to ours is ABC-Verify \cite{zen2021abc}, which also utilizes a blockchain and machine learning approach. However, they use the machine learning approach for news content, whereas we use a classifier for users' behavior. 

\section{Modeling the Misinformation Process}

The Oxford English dictionary defines misinformation as ``the act of giving wrong information about something''. Online misinformation, in this case, defines a form of misinformation that is transmitted through online tools such as websites and social networks. In various studies, disinformation is considered as a subset of misinformation, while some studies separate disinformation as the intentional form of misinformation \cite{zhou2020survey}. Although studies that try to distinguish disinformation and misinformation may yield the original intent of the spreader, in our model, we emphasize that there does not exist a practical difference affecting its propagation, since the intent of the spreader will be a hidden variable.

We define misinformation as a framework between senders and receivers on a platform that enables a publish/subscribe social network setting. This is the most generic form of a social network. We identify three intuitive mechanisms; Message Preparation Mechanism, Medium, and Message Propagation Mechanism. We further emphasize various latent variables that emerge with the introduction of Interpersonal Deception Theory. We argue that vulnerabilities in the mentioned mechanisms and variables worsen the spread of misinformation; hence, solutions that aim to stop misinformation should consider targeting these. 

\subsection{Interpersonal Deception Theory}

\begin{figure*}[tb]
      \centering
        \includegraphics[width=1\textwidth]{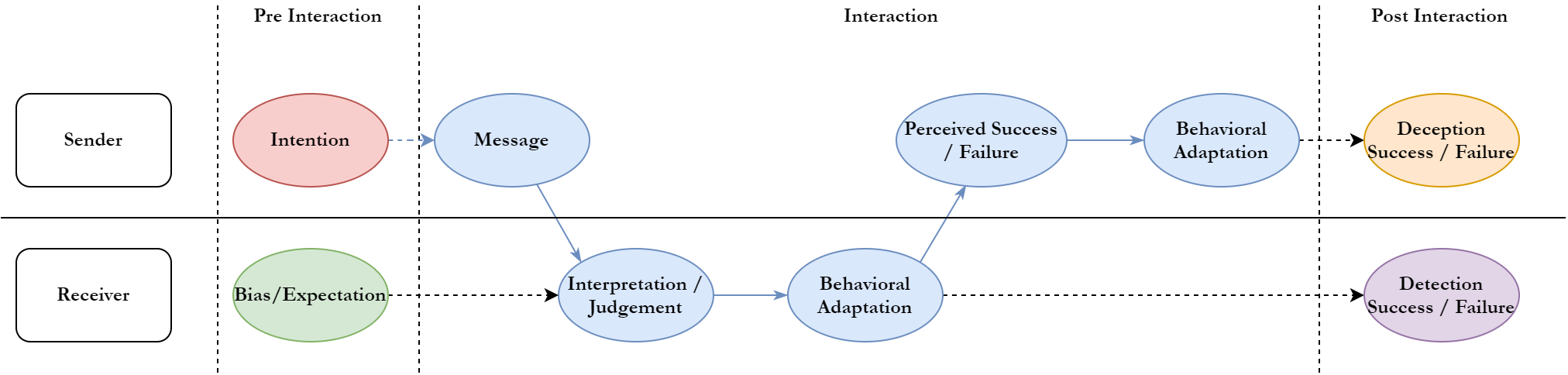}
        \caption{An adapted diagram of Interpersonal Deception Theory}
        \label{idt}
    \end{figure*}

Interpersonal Deception Theory (IDT) was established by Buller and Burgoon \cite{buller1996interpersonal}. It uncovers the dynamics of deception in interpersonal communication. According to the theory, communication occurs between a sender and a receiver in a repeated fashion based on some communication variables and sender/receiver properties which can affect the success of deception or detection. It is basically about face-to-face communication between individuals. A diagram of communication-based on IDT is presented in Figure \ref{idt}.

  \begin{figure*}[htb]
        \centering
        \includegraphics[width=1\textwidth]{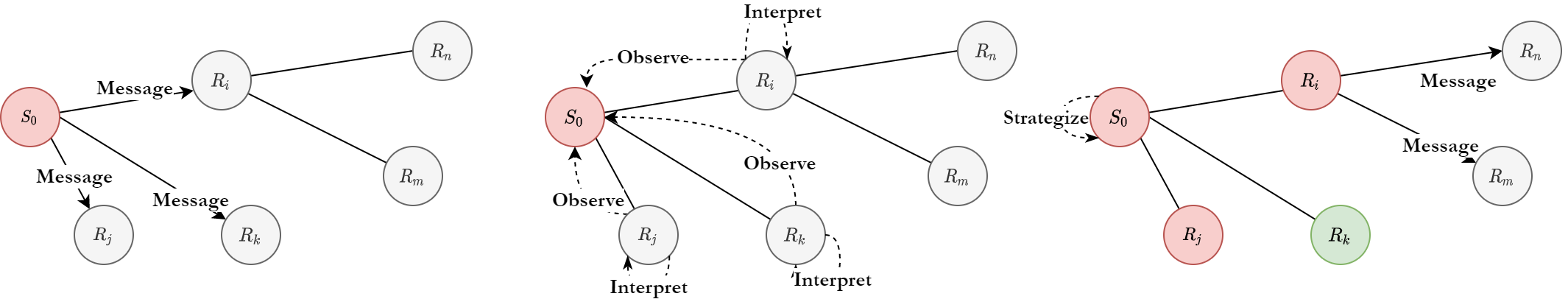}
        \caption{A networked model of misinformation propagation}
        \label{networkedmodel}
    \end{figure*}

In online communication, face-to-face clues are non-existent, but linguistic clues and cognitive dynamics are still in play, and they happen in a networked setting, which could potentially augment or affect communication dynamics. We adopt the misinformation model depicted in Figure \ref{networkedmodel}. In a scenario where people follow each other and there are mechanisms for sending messages and sharing content (e.g., Tweets, Retweet on Twitter, Sharing other platforms), misinformation can be sent via a message from an original sender who may or may not have malicious intent to influence other people in its proximity, which could, in turn, be adopted by its followers and spread to their neighbors. According to the IDT, the original sender with malicious intent can observe the reactions (via e.g., Share, Retweet, Like, Dislike, etc.) to adjust its behavior for future messages (transactions). A person with no malicious intent can also help spread the misinformation and indirectly serve the malicious intent, where the content of the message would not yield the original intent. Although people are rational beings and say that they can identify deception easily, the research argues against it \cite{bond2006accuracy} with only 54\% can identify deception, which is slightly better than tossing a coin. In addition, before even seeing the message, people would already have a perception of the subject of the topic; hence, this should also be taken into consideration \cite{zhou2020survey}.

However, a more generic model is needed to represent and identify certain parts of online communication that play a critical role in the dissemination or suppression of misinformation. We first explain each part and associated vulnerabilities.

 \begin{figure*}[tb]
      \centering
        \includegraphics[width=\textwidth]{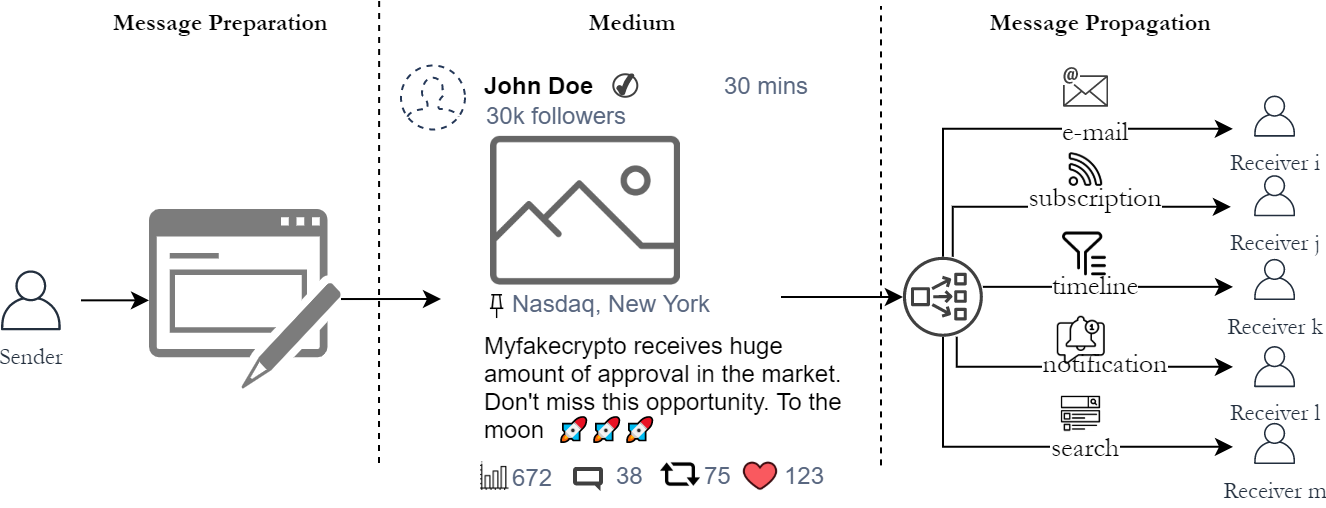}
        \caption{A generic model of online misinformation}
        \label{genericmodel}
    \end{figure*}

\subsection{Message Preparation Mechanism}

This part represents the initial part of the misinformation spread. A message is created by the sender and submitted in this stage. The online platforms generally enable this with a submission form, which could be decorated by the platform with various features such as location, tags of people, etc. This part is vital to stop misinformation even before it is introduced. This part is under the control of the online platform. The features of the message and the sender could hint at whether misinformation is present. 

\subsubsection*{\textbf{Message}}

A message can be generalized as a set of propositions. The propositions are either direct or indirect which are achieved by the lexical, syntactic, and semantic features aided with various other features such as videos, images, and other digital content. Once the message is delivered, it is interpreted by the receiver, not necessarily yielding the same propositions as they are created. The intent, however, is an intrinsic variable that could help shape the message and explain the rationale for the message being created. However, the intent is not present in the message since a person with no malicious intent could unintentionally spread a message with malicious intent, achieving the same results, even worse \cite{zhou2020survey, zhou2022fake}. A message can be described as a set of one or more propositions that are communicated via textual or other digital content. While it is possible to argue that each of these propositions should be true for the message to be true, it is still possible that even that is not enough.

Other methods may include fuzzy logic-based approaches where the message has a degree of its truthfulness. 

In this stage, various precautions could be implemented to prevent misinformation even before it is created, for instance, detecting a possibility of misinformation by analyzing the content of the message using various machine learning techniques. Recent studies are summarized in \cite{sahoo2021multiple}.

\subsubsection*{\textbf{Sender}}

There are various features such as previous history, profile, and the tendency of the sender that could indicate the existence of misinformation. One of the most hidden features of the sender is the intention \cite{zhou2022fake}, which is hard to tell. For instance, while the original poster of a fake story most likely has the intention to deceive, some other users may simply believe the original post and re-post the story. Possible vulnerabilities include weaknesses in the authorization and authentication of the user, such as the creation of Sybil or bot accounts \cite{alharbi2021social}. 

\subsection{Medium}
The medium is the basic functionality of the online platform. It connects people and allows interaction with posts with various means (e.g., likes, shares, retweets). It enables the people in the network to observe the statistics about the posts of other people or their own. It is when you see how many people watched a video or added a comment to it. This part is also managed by the online platform. Every textual, visual, and auditory clue provided by the medium could potentially augment and diminish the effects of misinformation attempts \cite{glenski2017rating}.

\subsection{Message Propagation Mechanism}

This part represents how the initial message is delivered to the other nodes using the medium of the platform. 

\subsubsection*{\textbf{Delivery}}

The delivery mechanism could be a simple publish-subscribe method with which subscribers are notified when a publisher publishes a message. However, the order that they receive the message could also be very important \cite{cohen2018exploring}. The order, again, is under the control of the online platform, which is proprietary in many cases. Online platforms employ various ranking algorithms to make some posts more visible and with various other external tools such as push notifications, newsletters, and e-mails. Transparency in this part cannot be guaranteed unless all the corresponding algorithms are made public and it is shown that there is no bias or temperament of visibility.

\subsubsection*{\textbf{Receiver}}
Most of the features of the Receiver are the same as the Sender but for the misinformation to be successful, the interpretation of the receiver is the most important factor. Expectations, knowledge level, and bias towards the message content or the sender are the features to consider. Failing to authenticate the validity of the receiver is also a problem for the dissemination of the information, as Sybil and Bot accounts can interact with a correct message in a malicious way. The reverse is also through where these accounts could be commanded to act in favor of an adversarial sender.

\section{A Generic Blockchain Approach}

\subsection{Background}

Blockchain is a distributed ledger technology that allows for the secure and transparent storage of data. It is a decentralized system, meaning that it is not controlled by any single entity, but rather is maintained by a network of participating nodes. Blockchain was developed as a way to enable secure and transparent online transactions without the need for a central authority or third-party intermediary.

Ethereum is a blockchain platform that was developed to enable the creation and execution of decentralized applications (dapps). These dapps are built using smart contracts, which are self-executing contracts with the terms of the agreement between buyer and seller being directly written into lines of code. The use of smart contracts allows for the automation of complex processes and can help reduce the costs and risks associated with traditional contracts.

Solidity is the programming language used to write smart contracts on the Ethereum platform. It is a high-level language, similar to JavaScript, that is designed to be easy to write and compile down to the low-level code that can be executed on the Ethereum Virtual Machine (EVM). Solidity allows developers to create smart contracts that are both functional and secure.

Ganache is a local Ethereum blockchain that can be used for testing and development purposes. It allows users to create a private blockchain where they can deploy and run smart contracts without having to connect to the main Ethereum network. This makes it a useful tool for experimenting with smart contract development and for testing the functionality of dapps before deploying them to the main network.

\subsection{Proposed Approach}
In social networks such as Twitter, Facebook, and Reddit, we see users creating posts and other users interacting with those posts through various means such as liking, retweeting, and sharing. Each of these platforms has its ways with which the posters accumulate reputation-like standing within the social network. This can either be implicit (e.g., number of followers, number of retweets) or explicit (e.g., Karma system in Reddit). Yet, there does not exist a metric to identify the validity and the reliability of the posts in terms of truthfulness, in addition to the lack of responsibility for the posters as well as the other users interacting with the posts. Therefore, the social platform needs to accommodate the validity, immutability, and non-repudiation of any transaction regarding posts to facilitate more misinformation-free communication.

A blockchain can deliver an environment of such transactions where validity, immutability, and non-repudiation are ensured while also making sure that no single entity determines how the transactions are handled. Let us define an environment where there exist users who are capable of posting stories and voting on stories posted by other users. Votes denote the personal judgment of each user regarding the truthfulness of the story. This environment aims to exploit the benefits of collective intelligence provided by crowdsourcing to identify the validity of the information.  

We implement the crowdsourcing mechanism using Ethereum-based smart contracts in the Solidity programming language. We use the Ganache framework to create, compile, deploy and test the smart contracts on a local development environment. Each user has a unique identifier on the blockchain. The smart contract enables users to post stories and vote on stories that are not their own. Both of these actions are called transactions and are immutably stored on the blockchain. Extra features of the users, such as their reputation, are also stored on the blockchain. 

Once a story is posted, users start upvoting (+1) or downvoting (-1) the story, indicating their judgment that the story is true or false (fake), respectively. For determining when to terminate voting for a particular story, we experimented with mechanisms such as calculating the momentum of the incoming votes, the entropy of the votes, the entropy change via Kullback-Leibler divergence, monitoring for a particular portion to vote on the story or simply monitoring for a maximum number of votes. However, we decided to use a combined approach to determine the equilibrium for a particular story that involves a threshold mechanism determined by a classifier, entropy, vote count, and Lyapunov exponents which we describe in detail in the following section (See Algorithm \ref{equi}). This is also the time we distribute the rewards and punishment in the blockchain involving this particular story.

\paragraph{\textbf{Lyapunov Exponents}}

Lyapunov exponents are often used in the study of chaotic systems, as they provide a quantitative measure of the degree of chaos in a system. They are also used in the study of complex systems, as they provide a way to understand the behavior of a system at a macroscopic level, even if the system is made up of many interacting parts.

In general, a positive Lyapunov exponent indicates that the system is chaotic and sensitive to initial conditions, while a negative Lyapunov exponent indicates that the system is stable and insensitive to initial conditions. A system with zero Lyapunov exponents is considered to be periodic. If some of the Lyapunov exponents are positive and some are negative, then the system is somewhere between stable and chaotic.

First, let us define the variables that we will be using:
\begin{itemize}
\item $\mathbf{x}_t$: The state of the system at time $t$.
\item $\mathbf{F}(\mathbf{x}_t, t)$: The dynamics of the system, which describes how the state of the system changes over time.
\item $\mathbf{V}_t$: The Jacobian matrix of the system at time $t$.
\item $\Delta t$: The time step size.
\end{itemize}

With these definitions, we can express the Lyapunov exponents as follows:

$$\mathbf{V}_{t+\Delta t} = \mathbf{V}_t + \frac{\partial \mathbf{F}}{\partial \mathbf{x}_t} \mathbf{V}_t \Delta t$$

This equation describes how the Jacobian matrix changes over time. To calculate the Lyapunov exponents, we can iterate this equation over a series of time steps, starting from an initial time $t_0$ and ending at a final time $t_1$.

$$\mathbf{V}_{t_0} = \mathbf{I}$$
$$\mathbf{V}_{t+\Delta t} = \mathbf{V}_t + \frac{\partial \mathbf{F}}{\partial \mathbf{x}_t} \mathbf{V}_t \Delta t$$
$$\mathbf{V}_{t+2\Delta t} = \mathbf{V}_{t+\Delta t} + \frac{\partial \mathbf{F}}{\partial \mathbf{x}_{t+\Delta t}} \mathbf{V}_{t+\Delta t} \Delta t$$
$$\vdots$$
$$\mathbf{V}_{t_1} = \mathbf{V}_{t_1-\Delta t} + \frac{\partial \mathbf{F}}{\partial \mathbf{x}_{t_1-\Delta t}} \mathbf{V}_{t_1-\Delta t} \Delta t$$

Once we have calculated the final Jacobian matrix $\mathbf{V}_{t_1}$, we can use it to calculate the Lyapunov exponents as follows:

$$\mathbf{L} = \text{eig}(\mathbf{V}_{t_1})$$

Here, $eig$ is a function that calculates the eigenvalues of a matrix, and $\mathbf{L}$ is the list of Lyapunov exponents.

\paragraph{\textbf{Shannon Entropy}}

\begin{equation}
H = -\sum_{i=1}^{n} p_i \log_2 p_i
\end{equation}

where $n$ is the number of different transactions, and $p_i$ is the probability of transaction $i$.

To calculate the probabilities $p_i$, we need to count the number of times each transaction occurs in the data. Let $N_i$ be the number of times transaction $i$ occurs in the data, and let $N$ be the total number of transactions. Then, the probability of transaction $i$ is   $p_i = \frac{N_i}{N}$.

\paragraph{\textbf{Vote Count}}
As both Lyapunov exponents and entropy work better with more data, we need to handle the case where there exist a low number of votes on a story. To this purpose, we add another metric called "vote count", which means that the number of votes should exceed a minimum amount $C_{min}$ to be able to terminate voting on a story.

\begin{algorithm}
\caption{Decide whether a system of transactions has reached equilibrium}
\begin{algorithmic}[1]
\Require $\mathbf{L}$: Lyapunov Exponents 
\Require $\mathbf{H}$: Shannon Entropy
\Require $\mathbf{\tau}$ Entropy Threshold
\Require $\mathbf{C}:$ Count of transactions (Votes on story)
\Require $\mathbf{C_{min}}$: Minimum number of transactions
\Procedure{Equilibrium}{$\mathbf{L, H, \tau, C, C_{min}}$}
  \State $LEquilibrium \gets True$
  \For{$\lambda$ in $\mathbf{L}$}
    \If{$\lambda \ge 0$}
      \State $LEquilibrium \gets False$
      \State \textbf{break}
    \EndIf
  \EndFor
  \If{$LEquilibrium$ and $H < \tau$ and $C>=C_{min}$}
    \State \textbf{return True} \Comment{"The story is stable and in equilibrium."}
  \Else
    \State \textbf{return False} \Comment{"The story is chaotic or unstable and not in equilibrium."}
  \EndIf
\EndProcedure
\end{algorithmic}
\label{equi}
\end{algorithm}

\subsection{Learning Problem}

  \begin{figure*}[!htb]
        \centering
        \includegraphics[width=1.05\textwidth]{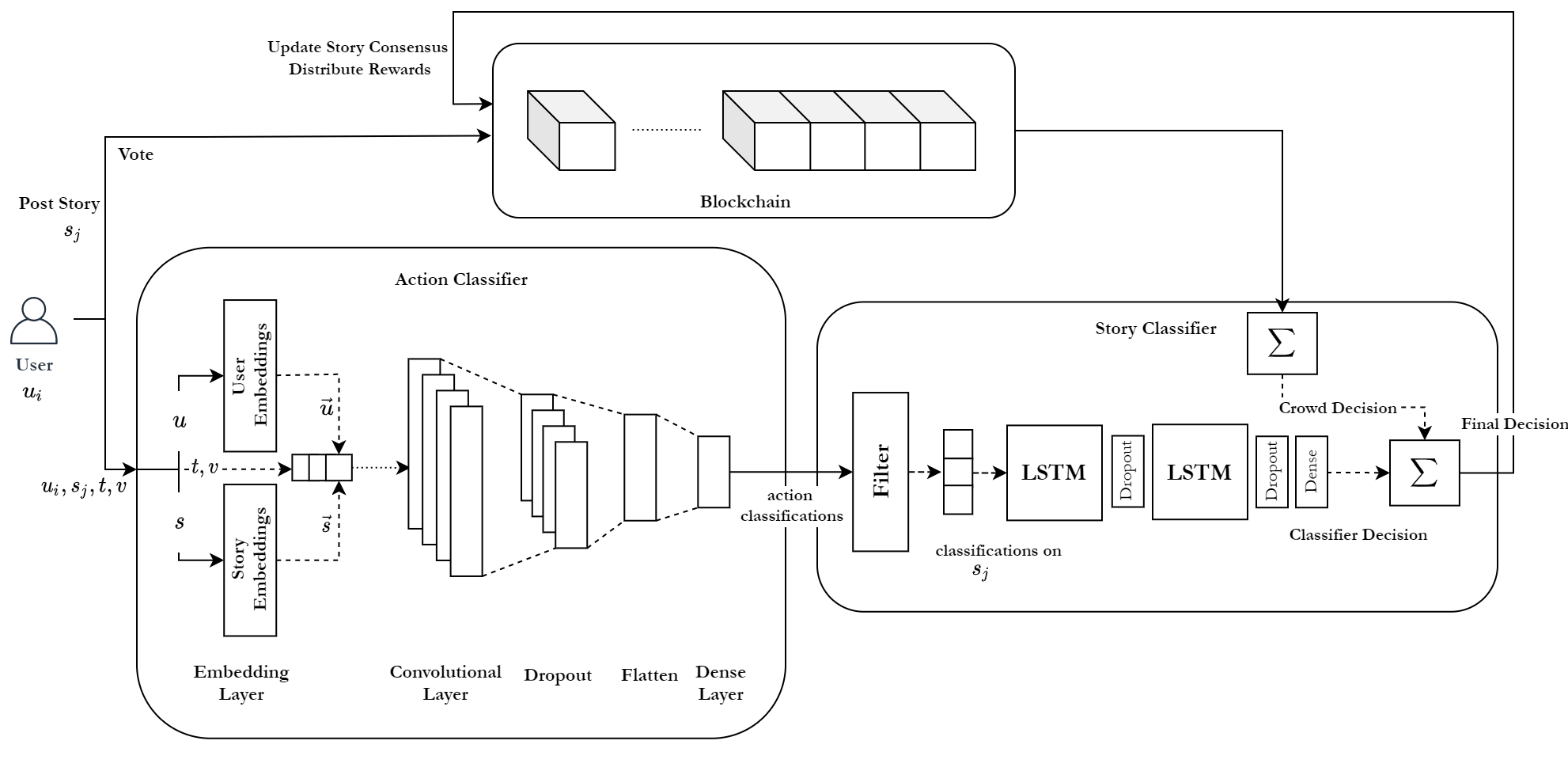}
        \caption{Overall Architecture of the Solution}
        \label{architecture}
    \end{figure*}

As we previously mentioned, a crowdsourcing system should be resilient against malicious behavior. This requires further support from sophisticated algorithms. While it is possible to create a rule-based behavior detection system, the complications would yield the need for a more automated system that is capable of learning over time and detecting more sophisticated adversarial scenarios that are unknown at the time. As there exists a rising concern about the use of bots in the literature \cite{ferrara2016rise, ross2019social}, the system should be more intelligent than those bots. To this purpose, we employ a two-stage learning architecture combining two different classifiers learning two different models, as depicted in Fig. \ref{architecture}. The first one - \textit{"Action Classifier"}, learns to distinguish malicious behavior. For this, it is fed time-series data consisting of quadruples in the form  (user, story, type, vote).

We consider two key aspects of representing users, stories, and user votes to solve the learning problem. The first aspect is the sequential form of the data, which captures patterns in user behavior. The second aspect is the need to maintain user-user and story-story similarities. To address these considerations, we propose representing actions as triples of (user, story, vote), which preserves the sequential dimension but fails to provide user-user or user-story similarity. Another approach is to represent actions as a user-story matrix with votes as values, which preserves general similarity but lacks the sequential dimension and suffers from sparsity. Ultimately, we propose combining both approaches to fully capture both aspects. We employ two separate embedding layers that learn representative vectors of users and stories of sizes $log_2 |U|$, $log_2 |S|$, respectively, where $|U|$, $|S|$ represent the maximum number of users and stories, respectively. We then concatenate those with their corresponding type and vote values. These are then fed into a convolutional layer that utilizes an additional max-pooling layer. A dropout layer is used to prevent overfitting, and finally, a dense layer outputs the classification on stories using a $tanh$ activation function.

To classify the story based on the previous behavior of the users and corresponding classifications of the \textit{Action Classifier}, we implement the \textit{Story Classifier} that utilizes two LSTM layers with a dropout layer after each and a final dense layer. The input data is the time-series classifications of the first classifier. Once we predict the label for the story, it is then combined with the crowd decision as the final decision. If this value is greater than a threshold, it is marked as consensus and the story is updated on the blockchain, and the voting is terminated. After this, a reward is distributed to users that correctly identify the consensus label, and punishment is distributed otherwise. The poster is also rewarded if the posted story is true, and punished otherwise. All the cumulative rewards and punishments constitute the reputation score of the user. The rewards and punishments are kept very straightforward; the user that correctly identifies a false story or a correct one receives 1 unit of reputation, otherwise punished by -2 units. The poster of a correct story is rewarded 2 units of reputation and punished by -4 units otherwise.

\section{Experimental Evaluation}

The experimental setup contains a customizable environment for various attack scenarios and a variable number of users and transactions (stories and votes). By varying the parameters, we measure and analyze the behavior of the system. 

We define users who can distinguish true and false stories as "normal". These users do not have any other agenda than making a good judgment.

We also define a user type called "trolls". These users downvote a true story and upvote a false one. Even though these do not have specific targets, they can be quite disruptive. This is sometimes called the "naive" attack.

In another type of attack, malicious users build a good reputation
over time, then use this reputation to perform short-lived sets of malicious actions \cite{sun2006trust}. This attack is called the "on-off" or the "oscillation" attack. The related users are called the "traitors" by \cite{MARTI2006472}, which we implemented in the system as well.

Random behavior is one of the hardest attacks to countermeasure in reputation systems \cite{hasan2022privacy}. Users with this behavior act randomly, regardless of the story. 

Slandering attack or badmouthing is the act of downvoting specific users but acting normally against others. The reverse of this is called "whitewashing" or "promoting", which is the act of upvoting certain users' posts even if they are fake. 

When there are groups of users involved in more sophisticated types of attacks, these are called "orchestrated" or "coalitional" attacks. There are many ways user groups can attack. In this study, we implemented "orchestrated slandering", in which a group of users slander a set of target users, and "orchestrated whitewashing", in which they upvote a set of target users. There can be many other attacks, such as one version of the oscillation attack but performed by groups of users. To summarize, there exist the following malicious user types in our simulations in addition to the non-malicious user type normal: troll, random, traitor, and, orchestrated. There are also target users for slandering and whitewashing who act as normal users.

\begin{figure*}[tb]
\centering
\subfigure[Train]{\includegraphics[width=0.49\textwidth]{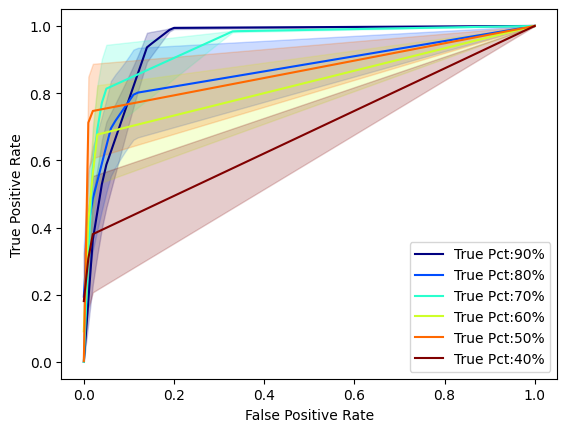}}
\subfigure[Test]{\includegraphics[width=0.49\textwidth]{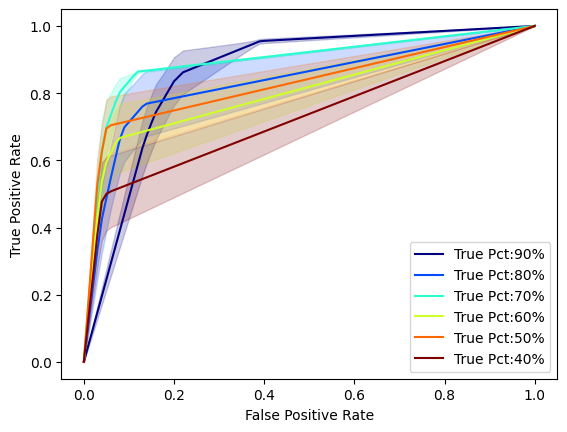}}

\caption{ROCs for train and test datasets where there exist  $n=100$ users, $k=500$ stories, $v=2500$ votes. Percentages of the user types: normal:70, troll:5, random:5, traitor:5, orchestrated:5, slandering targets:5, whitewash targets:5}
\label{roc}
\end{figure*}

In our simulations, we use varying ratios of true and false stories. To understand the effect of initial story distribution on the final outcome of misinformation detection, we perform various tests. By keeping the number of users, the ratio for various types of users, and the number of stories the same, we perform a test only by varying the true/false ratio of stories. Fig. \ref{roc} gives the receiver operator curves (ROCs) along with their 95\% confidence intervals, as we performed the test 5 times for each. According to the curves on the train and test sets, we can observe that the classification performance is affected by the initial distribution, especially the true positive rate. 

To understand how the attack types affect the classification success of the system, we utilize Ordinary Least Square (OLS) Regression analysis by using Precision, Recall, F1, and Accuracy values from 1000 runs as our dependent variables. In the tests, the percentage of normal users ranges from 30 to 70, while the remaining is distributed to the malicious types where there exist at least 5 percent of each type. The results are given in Table \ref{OLS}. If we look at the coefficients, the normal type of users contributes positively to the metrics, as expected, and the other types of users contribute negatively. We can generalize that troll attacks have the largest effect on the outcome negatively, followed by traitors, random and orchestrated, in this particular order. This is a surprising finding because trolling is not a particularly sophisticated adversarial strategy.

\begin{table}[tb]
\caption{OLS regression results for various attack types: \textit{1000 runs, normal type 30-70\%, other types distributed and at least $>$5\%}}
\centering
\resizebox{\textwidth}{!}{\begin{tabular}{|rrrrr|rrrr|}
\hline
 
 &   \multicolumn{4}{c|}{\hfill Precision \hfill\textit{$R^2 = 0.621 $}} &
      \multicolumn{4}{c|}{\hfill Recall \hfill \textit{$R^2 = 0.705 $}} \\
\hline
Type & Coef & Std Err & t  & $P > |t|$ & Coef & Std Err & t  & $P > |t|$ \\

\hline
normal&  -2.945e-05	&7.02e-05& -0.419& 0.675&  0.0003& 0.000& 2.315& 0.021\\
troll&  -0.0044& 0.000&-37.270& 0.000&  -0.0097& 0.000&-42.495& 0.000  \\
random& -0.0018& 0.000&-15.585& 0.000&  -0.0044& 0.000&-20.085& 0.000  \\
traitor&-0.0033& 0.000&-29.777& 0.000&-0.0074& 0.000&-34.168& 0.000\\
orchestrated&-0.0002& 0.000& -1.873& 0.061&-0.0002& 0.000& -0.712& 0.476\\
\hline
& \multicolumn{4}{c|}{\hfill F1 \hfill \textit{$R^2 = 0.736 $}} &
      \multicolumn{4}{c|}{\hfill Accuracy \hfill \textit{$R^2 = 0.705 $}} \\
\hline
normal&  0.0002&9.86e-05&1.630&0.103	& 0.0003& 0.000& 2.315& 0.021 \\
troll&-0.0078&0.000& -46.832&0.000& -0.0097& 0.000&-42.495& 0.000 \\ 
random& -0.0034&0.000& -21.230&0.000& -0.0044& 0.000&-20.085& 0.000\\
traitor&-0.0059&0.000& -37.403&0.000& -0.0074& 0.000&-34.168& 0.000 \\
orchestrated& -0.0002&0.000&  -1.117&0.264& -0.0002& 0.000& -0.712& 0.476\\ 
\hline
\end{tabular}}
\label{OLS}
\end{table}

While most of the results are significant, the effect of orchestrated attacks is not shown to be p<0.05, which means either we have to run more tests or the attack actually does not have a significant impact on the overall outcome, while benefiting the attackers. It should be noted that $R^2$ values also show relatively good models. 

\begin{figure*}[htb]
\centering
        \includegraphics[width=1.00\textwidth]{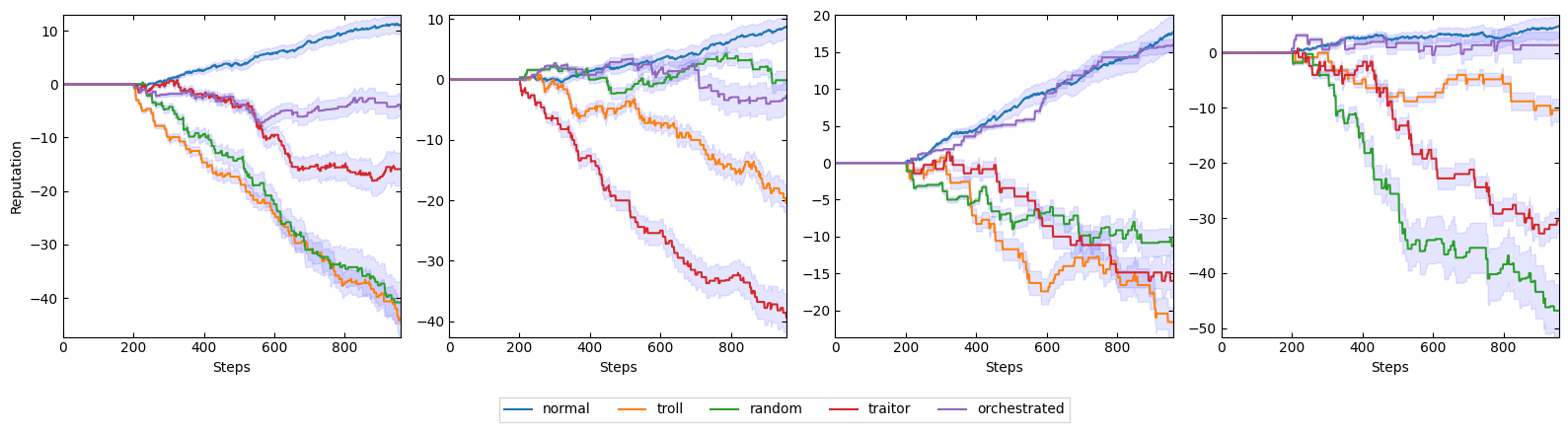}
        \caption{Reputation means of user types over time, $n=100, s=200, v=1000$}
        \label{rep-means-1}
\end{figure*}

\begin{figure*}[htb]
\centering
        \includegraphics[width=1.00\textwidth]{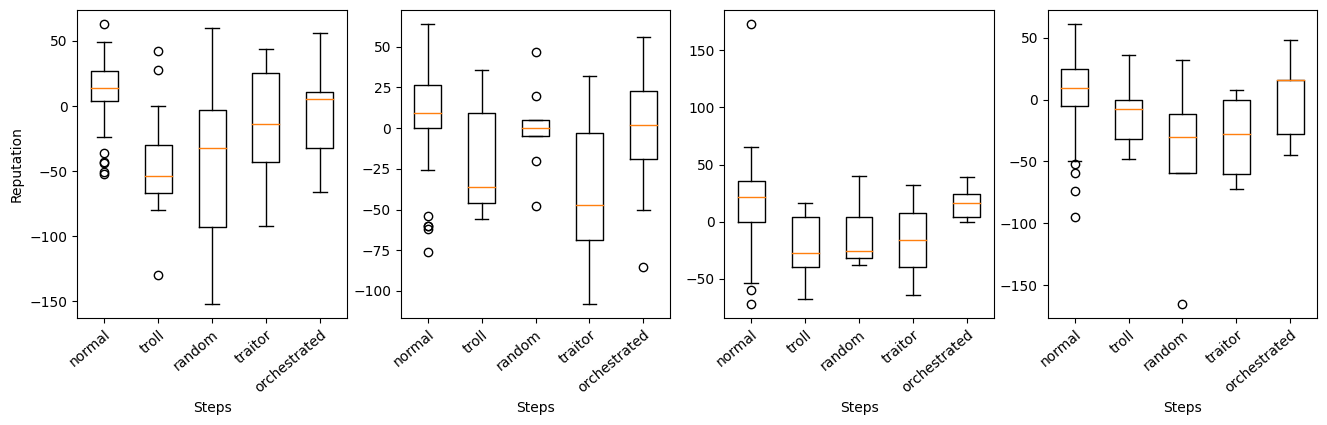}
        \caption{Final Reputation Distributions of User Types, $n=100, s=200, v=1000$}
        \label{rep-dist-1}
\end{figure*}

\begin{figure*}[htb]
\centering
        \includegraphics[width=1.00\textwidth]{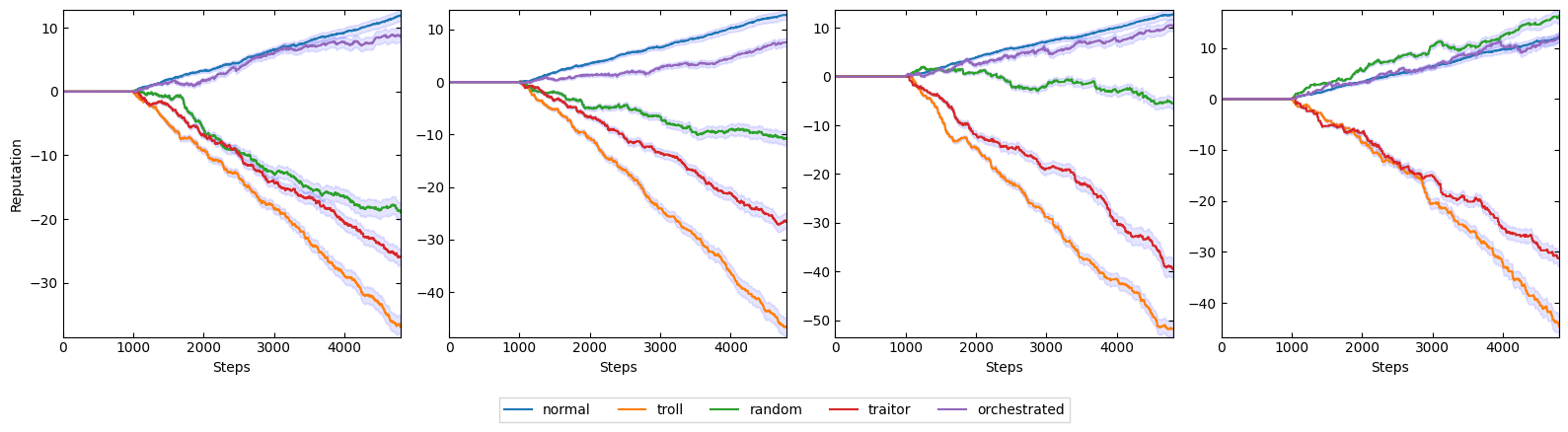}
        \caption{Reputation means of user types over time, $n=500, s=1000, v=5000$}
        \label{rep-means-2}
\end{figure*}

\begin{figure*}[htb]
\centering
        \includegraphics[width=1.00\textwidth]{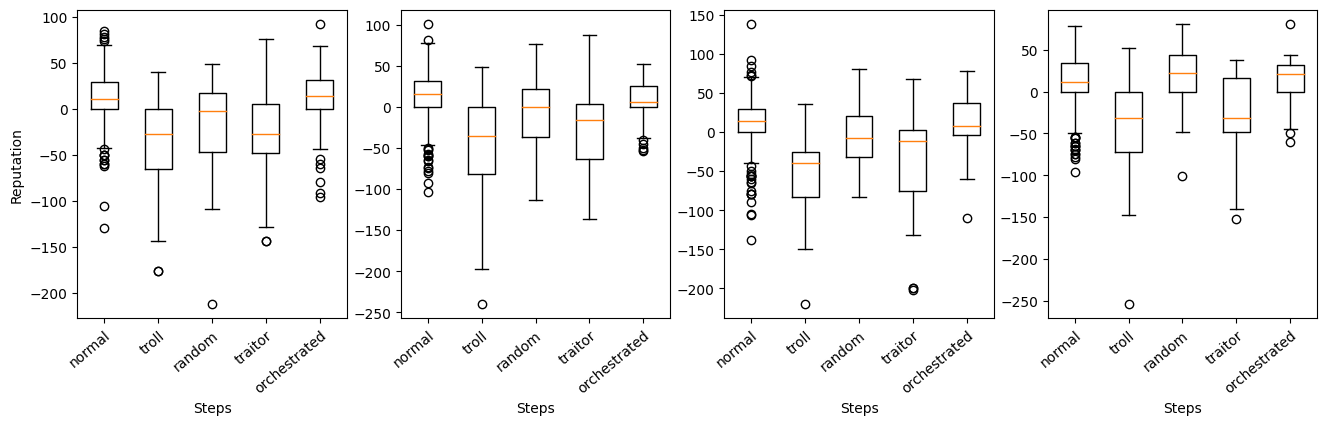}
        \caption{Final Reputation Distributions of User Types, $n=500, s=1000, v=5000$}
        \label{rep-dist-2}
\end{figure*}

\begin{figure*}[htb]
\centering
        \includegraphics[width=1.00\textwidth]{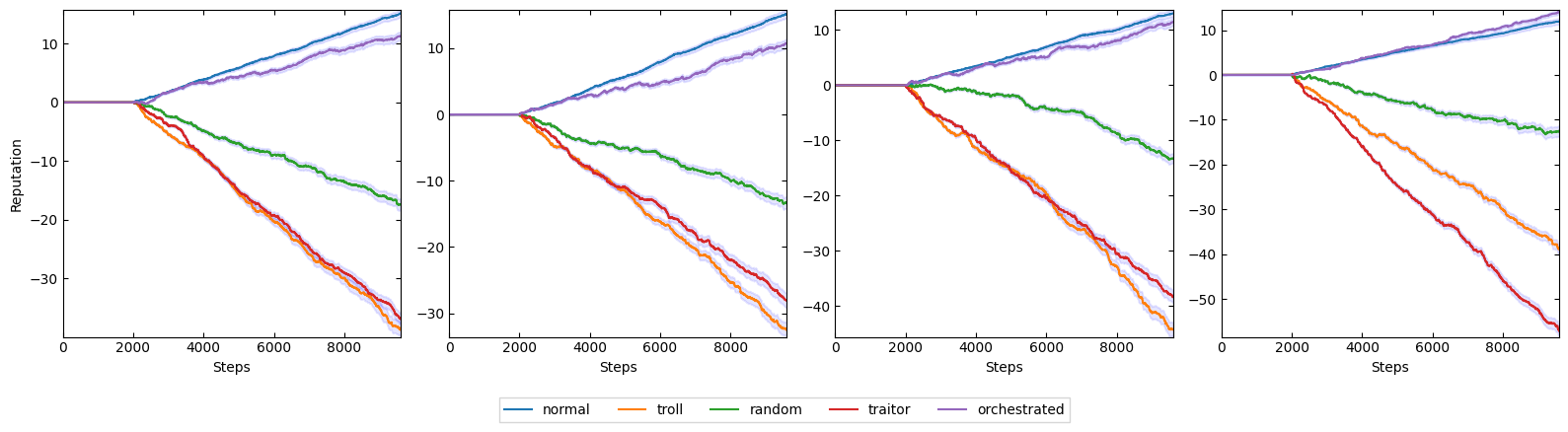}
        \caption{Reputation means of user types over time, $n=1000, s=2000, v=10000$}
        \label{rep-means-3}
\end{figure*}

\begin{figure*}[htb]
\centering
        \includegraphics[width=1.00\textwidth]{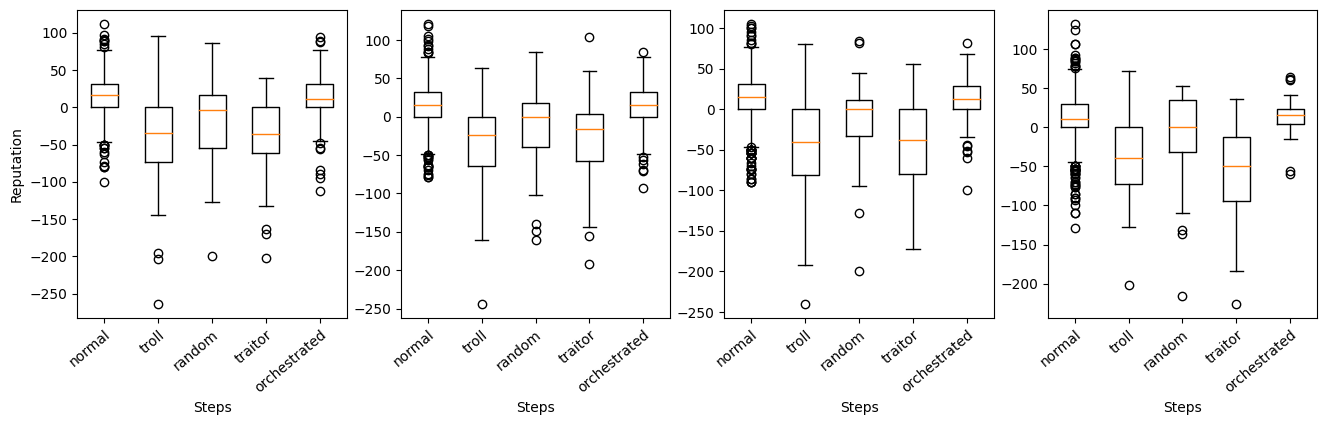}
        \caption{Final Reputation Distributions of User Types, $n=1000, s=2000, v=10000$}
        \label{rep-dist-3}
\end{figure*}

To understand how the reputation changes over time and how it is distributed for various user types, we perform various experiments. In these experiments, we set up various User-Story-Count triples and vary the percentage of initial user type distribution. In these experiments, the normal user type varies from 42\% to 70\%, target user types are fixed at 5\%, and malicious user types are equally distributed. For instance, for 100 users, 200 stories, and 1000 votes, Fig. \ref{rep-means-1} gives the results for varying ratios of user types. Fig. \ref{rep-dist-1} shows the final distributions of reputations for these scenarios. Fig. \ref{rep-means-2} and Fig. \ref{rep-dist-2} show the results for 500 users, 1000 stories, and 2000 votes, while Fig. \ref{rep-means-3} and Fig. \ref{rep-dist-3} present the results for 1000 users, 2000 stories, and 10000 votes.

\begin{table}[htb]
\caption{Precision, Recall, F1, and Accuracy Metrics for 3 different user-story-vote (U-S-V) distribution: Dist 1: 100 users 200 stories 1000 votes; Dist 2: 500 users 1000 stories, 5000 votes; Dist 3: 1000 users, 2000 stories, 10000 votes. The user type distribution per U-S-V distribution: 42-12-12-12-12-targets, 50-10-10-10-10-targets, 62-7-7-7-7-targets, 70-5-5-5-5-targets. }
\centering
\resizebox{\textwidth}{!}{
\begin{tabular}{|r|rrrr| rrrr|}
\hline

 & \multicolumn{4}{c|}{Metrics} & \multicolumn{4}{c|}{Malicious Detection} \\
Run & Precision & Recall & F1 & Accuracy & Troll & Random & Traitor & Orchestrated\\
\hline
\multirow{4}{*}{\raggedright U-S-V Dist. 1}&  0.81 & 0.68 & 0.74 & 0.68 & 0.96 & 0.96 & 0.85 & 0.33 \\
& 0.86 & 0.79 & 0.82 & 0.79 & 1.00 & 0.96 & 0.68 & 0.00 \\
& 0.89 & 0.83 & 0.86 & 0.83 & 0.82 & 0.8 & 0.83 & 1.00 \\
& 0.91 & 0.91 & 0.91 & 0.91 & 0.80 & 0.67 & 0.57 & 0.00\\ 
\hline
\multirow{4}{*}{\raggedleft U-S-V Dist. 2}& 0.83 & 0.70 & 0.76 & 0.70 & 0.90 & 0.94 & 0.92 & 0.57\\
& 0.85 & 0.74 & 0.79 & 0.75 & 0.89 & 0.93 & 0.83 & 0.18\\
& 0.89 & 0.84 & 0.87 & 0.84 & 0.83 & 0.86 & 0.76 & 0.54 \\
& 0.92 & 0.90 & 0.91 & 0.90 & 0.91 & 0.88 & 0.82 & 0.10\\

\hline
\multirow{4}{*}{\raggedleft U-S-V Dist. 3}& 0.84 & 0.70 & 0.76 & 0.70 & 0.90 & 0.84 & 0.88 & 0.42\\
& 0.87 & 0.77 & 0.82 & 0.77 & 0.84 & 0.92 & 0.89 & 0.25\\
& 0.89 & 0.84 & 0.86 & 0.84 & 0.81 & 0.85 & 0.85 & 0.50\\
& 0.91 & 0.88 & 0.89 & 0.88 & 0.91 & 0.88 & 0.84 & 0.17\\
\hline
\end{tabular}}
\label{dist_metrics}
\end{table}

We provide the Precision, Recall, F1, and Accuracy metrics as well as the ratio of malicious detection of various attacks in Table \ref{dist_metrics}. The results indicate that the system can identify the attacks successfully, but both detection and management of reputation for the orchestrated type attacks require more work. The malicious user types have less reputation overall than the non-malicious (normal) ones, and the system performs as described in the results under malicious behavior.

\subsection{Twitter Case Study}
Birdwatch is a new feature that is being developed by Twitter as a way to combat misinformation on the platform. The idea behind Birdwatch is to allow users to contribute to a community-driven approach to addressing misinformation. Rather than relying solely on Twitter's algorithms and moderators to identify and flag misinformation, Birdwatch allows users to identify and add notes to tweets that they believe are misleading or false. These notes will be visible to other users, providing additional context and information that can help users make more informed decisions about the content they see on Twitter.

Birdwatch is currently in the early stages of development and is being tested in a limited pilot program. It is not yet available to all users on the platform. Twitter has emphasized that the goal of Birdwatch is to create a system that is transparent, open, and accountable, and that allows for a diverse range of perspectives and voices. The company has also stated that it will use a variety of mechanisms to prevent abuse and manipulation of the system. It is not yet clear when or how Birdwatch will be rolled out to the broader Twitter community.

In \cite{mujumdar2021hawkeye}, the authors investigate Birdwatch and show that it is vulnerable to adversarial activity. They propose different metrics as criteria that need to be satisfied for the quality of tweets, users, and notes. They annotate a dataset of 500 tweets included in the Birdwatch data from January - April 2021 and show that their reputation system "HawkEye" outperforms Birdwatch in classifying tweets accurately. 

To understand how the system described in our work performs against a human-curated dataset, we adopt the dataset provided in the study of "Hawkeye". We treat notes on tweets and the ratings on notes as votes by counting agreeing ratings on notes as the same value as their corresponding note and disagreeing notes as the opposite value of the note. We observe that there exist 2.8\% duplicates in the original data set and eliminate them, and perform the simulation on this new set. In this dataset, we are unable to identify malicious actions as we are unsure of the intent of the users. We denote only the false-story posting as malicious. Notwithstanding, our system performs as accurately and in some cases better than Hawkeye, considering the duplicate elimination and the time-series approach in our work compared to Hawkeye, which used the entire dataset and cross-validation. The results are provided in Table \ref {twitter}. We give the results of the train and test sets, and list Hawkeye's results from their original paper. We employ 80\%-20\% train and test split, where the train set is the earliest portion of the dataset.

\begin{table}[tb]
\caption{Comparison of Hawkeye and our system on the Twitter Birdwatch dataset}
\centering
\begin{tabular}{|l|c|c|c|c|}
\hline
Metric & \parbox[t]{2.3cm}{\raggedleft HawkEye \textit{Supervised}}  & \parbox[t]{2.5cm}{\raggedleft  HawkEye \textit{Unsupervised}}& \parbox[t]{2cm}{\raggedleft  Our Work  \textit{Train}} & \parbox[t]{2cm}{\raggedleft Our Work \textit{Test}}\\
\hline
Precision & 85 & 79 & 85 & 85 \\
Recall & 74 & 78 &  77 & 75 \\
F1 & 76 & 78 & 78  & 77 \\
\hline
\end{tabular}
\label{twitter}
\end{table}

\section{Discussion}
In this section, we discuss some important aspects regarding the deployment and success of blockchain and machine learning-based solutions proposed to effectively deal with the misinformation problem.

First of all, the positioning of the blockchain is quite important. While the distributed nature of the blockchain provides fault tolerance, it is only possible for such a blockchain to be run without any specific owner only if there exists an incentive-based mechanism to run a blockchain node. Additionally, the blockchain should provide an API for the integration of existing social networks or news-providing services. Such a scenario would possibly include the creation of a user in the blockchain with a public key and a wallet, that is pointed by a user profile in an existing social network. Each of the stories posted on the social network would have to be inserted into the blockchain as well. 

With users, stories, and related votes each having a representation in the blockchain would create possibilities to solve existing research problems as well since the existing network topologies would be flattened out in the blockchain, hinting at the bypassing of the computationally difficult problems.  

For instance, if the blockchain should have the possibility of storing which user has seen a story, we would solve the problem of determining which users were affected by a post. We would simply filter the database, instead of using various algorithms to traverse the network. The blockchain would have such an API \texttt{AddUserStorySeen (userid, storyid)}, and the social network would call this API when a user visits a story.

It should also be noted that if the blockchain would be able to serve multiple social networks and news-sharing platforms, then this feature could be used to reduce the effect of the same information being copied and spread over different domains. In addition, we would have a better opportunity to identify the originality of the posts \cite{huckle2017fake}. 

On the other hand, the identification and prevention of Sybil accounts and solving the multiple identities have not been solved in social networks without trading off the privacy issue. Solving this problem in a blockchain-based solution, such as the one in this work, is of utmost importance in addition to the anonymity problem. 

One other issue that needs to be considered is to also have the machine learning approach distributed to solve the misinformation problem. Having a single authority host, the classifier would jeopardize the transparency we describe in this work. The technical solution would require hosting the model on different nodes and verifying model correctness. 

In this work, we used two different classifiers. The main advantage of this is that we can teach the first classifier different attack forms and this would increase the explainability of the classifier compared to having a single box. 

It should also be noted that we have implemented known attacks in a limited setting. As a future work, we consider using reinforcement learning to generate different attacks, analyze and try to encounter them.

\section{Conclusion}

 In this paper, we identified the critical factors in deception by referencing the Interpersonal Deception Theory and provided a model for misinformation in online social networks. We described how blockchains are one of the candidates that can provide a better solution for the misinformation problem in terms of transparency, decentralization, validity, and immutability. We proposed and implemented a crowdsourcing mechanism on the blockchain, specifically on Ethereum, using smart contracts towards identifying true and false stories. We added a deep learning system that took part in identifying the malicious actions in the crowdsourcing mechanism and the final decision on the truthfulness of stories. We trained and tested the system under various types of attacks and evaluated the experimental results. We provided a case study on Twitter Birdwatch data and compared our results with a previous work on the same data set. We obtained promising results which suggest further studies involving the solution.  

\begin{acks}
None.
\end{acks}

\bibliographystyle{ACM-Reference-Format}
\bibliography{main}

\end{document}